# CHARACTERIZATION OF BASE ROUGHNESS FOR GRANULAR CHUTE FLOWS

Lu Jing and C.Y. Kwok*

*Department of Civil Engineering, The University of Hong Kong, Haking Wong Building, Pokfulam Road, Hong Kong*

Y.F. Leung

*Department of Civil & Environmental Engineering, The Hong Kong Polytechnic University, Hong Kong*

Y.D. Sobral

*Departamento de Matemática, Universidade de Brasília, Campus Universitário Darcy Ribeiro, 70910-900 Brasília, DF, Brazil*

\* corresponding author: fiona.kwok@hku.hk

**ABSTRACT**

Base roughness plays an important role to the dynamics of granular flows but is yet poorly understood due to the difficulty of its quantification. For a bumpy base made by spheres, at least two factors should be considered to characterize its geometric roughness, namely the size ratio of base- to flow- particles and the packing of base particles. In this paper, we propose a definition of base roughness, $R_a$, which is a function of both the size ratio and the packing arrangement of base particles. The function is generalized for random and regular packing of multi-layered spheres, where the range of possible values of $R_a$ is studied, along with the optimal values to create maximum base roughness. The new definition is applied to granular flows down chute in both two- and three-dimensional configurations. It is proven to be a good indicator of slip condition, and a transition occurs from slip to non-slip condition as $R_a$ increases. Critical values of $R_a$ are identified for the construction of a non-slip base. The effects of contact parameters on base velocity are studied, and it is shown that while the coefficient of friction is less influential, normal damping has more profound effect on base velocity at lower values of $R_a$. The application of present definition to other base geometries is also discussed.

# I. BACKGROUND

The dense flow of granular materials has been studied in various configurations, including plane shear, annular shear, chute flows, heap flows, and surface avalanches in rotating drums [1–3]. These configurations result in either confined flows with one or more moving boundaries, or free surface flows supported by at least one substrate/base. Extensive physical and numerical studies have illustrated the profound impact of these boundary conditions to the overall flow behavior in both two-dimensional (2D) and three-dimensional (3D) situations [2,4–21]. For instance, Pouliquen [8] found that base roughness is the key to the minimum thickness necessary to sustain a steady flow at a given inclination, and proposed a scaling law correlating this thickness with the mean flow velocity.

Unless otherwise required (e.g., frictionless side walls in [2,22,23]), non-slip condition is presumed in many studies [1,3,8,12,24–37]. This imposes zero velocity at the stationary base in chute flows and prohibits sliding at the moving walls in plane/annular shear flows. There are several arguments supporting the imposition of non-slip conditions. In industries, a sufficiently rough base (e.g., a conveyor belt) can maximize the transport of materials and facilitate the control of flow type [10]. By contrast, the failure of persisting non-slip condition may give rise to poorly developed shear flows, such as plug flows on frictional but planar bases [10], and inefficient energy transfer from the rotating apparatus to the flowing granular materials [20]. From the theoretical point of view, non-slip is the simplest scenario in solving the boundary value problem of granular flow modeling. It is also the most common case in geophysical situations, thus widely adopted in the mathematical models of landslides [34,38], debris flows [39] and the segregation in shallow granular avalanches [29,40].

To achieve the non-slip base condition in physical experiments, a rough substratum is usually constructed by gluing a layer of randomly packed particles [1,8,20,27,30]. The validity of non-slip conditions may be influenced by the packing density of such particles, and deviations of their grain sizes that may occur during the manufacturing process. In numerical simulations, a layer of equal-sized spheres can be fixed beneath the flow to serve as the bumpy base [2,10,24,31,36,41,42]. Several arrangements of base particles have been studied [10] and it is found that neither quasi-ordered nor perfect-ordered bases can ensure non-slip condition. In contrast, a sufficiently rough base can be obtained with random packing, which is associated with irregular bumps and spaces. In some studies of mono-disperse flows, the size and distribution of base particles are set to be identical with a random layer of the flowing particles [24,41]. Alternatively, one can increase base roughness by fixing larger particles on the substrate [11,15–18,20].

Despite these empirical instructions on the construction of a rough base, the quantitative representation of base roughness remains rarely discussed. Inspiring work includes [5], where a mathematic model is developed for the 2D motion of a single bead on a rough inclined line, and [11], which theoretically investigates the angle of stability of a single particle on a rough plane. In both studies, different base construction (i.e. spacing or packing) and size ratio between base particles and bulk particles are considered. These indicate the possibility of defining base roughness as a function of size ratio and spacing/packing. Such a definition is particularly useful for better understanding on the boundary effect in granular flows (e.g. [5,6,8,10,11,17,20]), and size segregation in bi-disperse flows down a rough incline [36,42,43]. In the latter case, one known issue is that size segregation will lead to variation of relative base roughness and thus the flow kinetics [36,42]. Crystallization may occur upon basal sliding, and the development of se-gregation may be affected [36]. Basal effect in chute flows is further illustrated in Sec. II, following a brief description of the studied scenario and the adopted numerical scheme for the current work. To characterize base roughness in a quantitative manner, a newly-defined indicator is presented in Sec. III, including its definition, generalization to multi-layer situations, and boundedness. Phase diagrams have been established using the new definition for both 2D and 3D configurations. These can predict the slip/non-slip basal condition at a wide range of inclinations, according to the size ratio and packing of particles. Section IV discusses a few practical key points regarding the application of this indicator, such as the effects of microscopic inter-particle friction and macroscopic geometric roughness. Extension of the proposed approach to different surface types (e.g. curved, circular) and shapes (triangle, hemisphere/semicircle) are presented at the end of Sec. IV.

## II. BASAL CONDITION IN CHUTE FLOWS

### A. Case setup

Chute flow is an experimental paradigm of natural landslides and avalanches (Fig. 1). As a flow of granular material is continuously fed through a gate with controlled opening, a steady uniform flow of the desired thickness is developed at a given inclination. The steady, fully-developed (SFD) state is reached when the flow height, mass flow rate and thus kinetic energy are not varying in the flow direction [2]. In SFD, the velocity profile typically obeys Bagnold's scaling and shear stress is proportional to the square of shear rate [8,24]. The side-wall effect is negligible if the chute is sufficiently wide, in which case the experimental setup can be simplified as the periodic samples adopted in many numerical simulations [24,31].

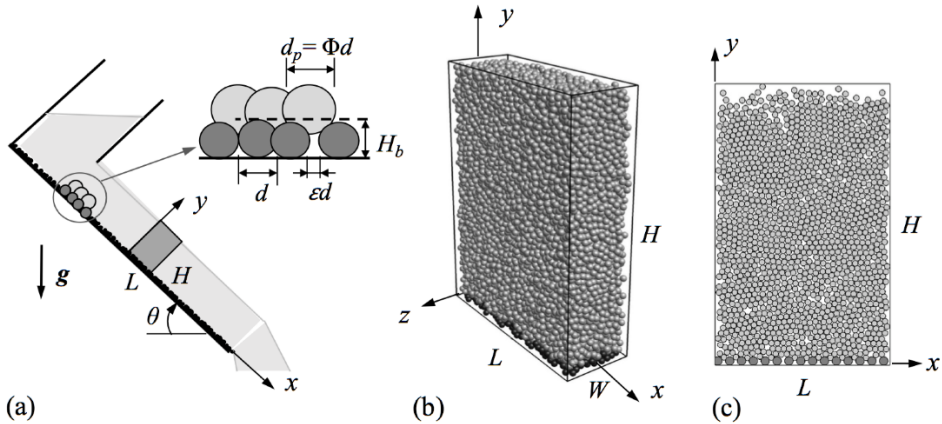

**FIG. 1.** Case setup. (a) chute flow in experiments, (b) 3D and (c) 2D periodic elements in DEM.

The presented numerical simulations are performed using Discrete Element Method (DEM). As shown in Fig. 1, periodic boundaries are imposed in the flow ($x$) and vorticity ($z$) directions, while a rough base is formed normal to $y$-direction. The sample is free of constraint at the top. Spherical particles are randomly poured into the sample box under gravity, and contact properties are tuned to achieve a close packing (packing density ~0.6). After the sample is generated, gravity is tilted to a designed inclination, $\theta$. The inclination is set close to the upper limit of SFD flows [2], i.e. $\theta = 30°$, as the cases aim mainly to examine the implementation of non-slip condition. In addition, several different inclinations are also presented in later sections for a more universal characterization of slip/non-slip condition.

The flow particles have diameter $d_p$ = 0.005 m, with sample length $L$ = $30d_p$, width $W$ = $10d_p$ and height $H$ = $40d_p$. The sample dimensions are chosen to strike a balance between eliminating boundary effects and maximizing computational efficiency. Mechanical properties of the particles include $\rho$ = 2500 kg/m$^3$, Young's Modulus $E$ = 5 GPa and Poisson's ratio $v$ = 0.35. The contact force is calculated using Hertz model [20,24], with normal damping given by $\gamma_n$ = ln$e$/$\Delta t$, where $e$ is the coefficient of restitution and $\Delta t$ the collision time. No tangential damping is considered. The tangential force is calculated following Coulomb friction criterion, with $|F_t| \leq |\mu F_n|$, where $\mu$ is the coefficient of friction, and $F_n$ and $F_t$ are normal and tangential contact forces, respectively. In the present study, the typical value of $\Delta t$ is $10^{-5}$, which ensures numerical stability. Since the focus is laid on the geometric roughness instead of mechanical properties, $e$ = 0.5, $\mu$ = 0.5 are consistently used for most simulations. Additional analyses are presented in Sec. IV to investigate the effects of varying $e$ and $\mu$ between the flow and base particles.

## B. Base generation

A variety of particle sizes and generation strategies are adopted (Table I). Main controlling variables are notated as follows: The size ratio between flow particles ($d_p$) and base particles ($d$) is denoted by $\Phi$ = $d_p/d$, while the (mean) spacing measured over a layer is notated $\varepsilon$ as [5]. Notations can be found in the magnified schematics in Fig. 1(a). Some equivalent expressions of $\varepsilon$ are the packing density of a layer, $\eta$, and compactness, $c$, which is the ratio of the area occupied by the projection of spheres within a layer to the surface area of the layer [11]. It can be derived that for a layer with all centroids coplanar, $\eta$ = (2/3)$c$. A base can comprise multiple layers, in which case the packing density of its surface is denoted by $\eta_s$. The total thickness of a base is $H_b$, which measures the height of the bounding box of all base particles.

**Table I. Major simulations performed.**

| Set | 2D/3D | $\Phi$ | $\theta$(°) | Base generation | Sketch |
|---|---|---|---|---|---|
| 1 | 3D | 0.4 | 30 | | - |
| 2 | 3D | 0.5 | 30 | | - |
| 3 | 3D | 0.67 | 30 | Random: $H_b/d$ = 1.0, 1.2, 1.4, 1.6, 1.7, 1.8, 1.9, 2.0 | - |
| 4 | 3D | 1.0 | 30 | | Fig. 2(a–c) |
| 5 | 3D | 1.25 | 30 | | - |
| 6 | 3D | 2.0 | 30 | | - |
| 7 | 3D | 1.0 | 30 | Random: One layer: $\eta$ = 0.2, 0.3, 0.35, 0.4, 0.5, 0.6; Two layers: $\eta_s$ = 0.2, 0.3 | Fig. 2(d–f) |
| 8 | 3D | 1.0 | 30 | Ordered: Dense ordered packing, optimal packing | Fig. 2(g–i) |
| 9 | 3D | - | 30 | Fictional flat plane | - |
| 10 | 3D | 0.5–2.0 | 20–28 | Random: $H_b/d$ = 1.0, 1.2, 1.4, 1.6, 1.7, 1.8, 1.9, 2.0 | - |
| 11 | 2D | 0.5–2.0 | 25 | Ordered: $\varepsilon$ = 0, 0.2, 0.4, 0.6, 0.8, 0.9, 1 | Fig. 2(j–l) |

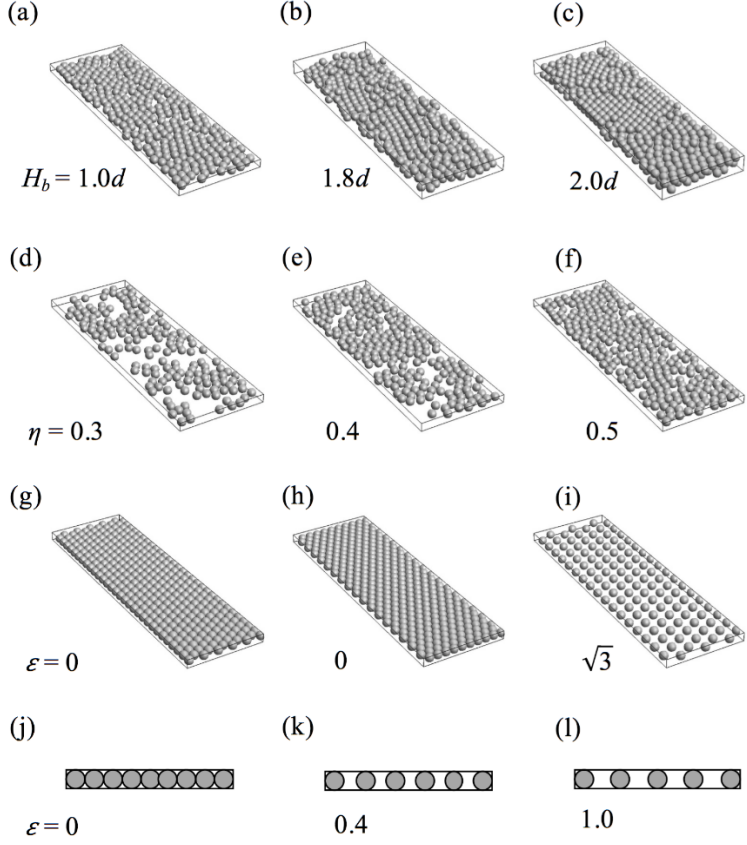

**FIG. 2.** Base generation. Random packing in 3D by (a)–(c) base thickness and (d)–(f) packing density; regular packing by spacing in (g)–(i) 3D and (j)–(l) 2D.

In Sets 1–6, different bases are generated by specifying $H_b = 1.0$–$2.0d$. If $y = 0$ is the designed top surface of a base, the generation procedure is by firstly placing a wall at $y = -H_b$, then pouring base particles onto the (bottom) wall, and finally trimming particles beyond $y = 0$. By varying $H_b$, surfaces with different distributions of bumps are produced on top of a dense layer (Fig. 2). In Set 7, a different strategy is used to generate the base for $\Phi = 1.0$, in which a layer of particles with different packing densities, $\eta = 0.2$–$0.6$, is randomly generated. By either approach (Sets 1–6 or Set 7), a random spacing is achieved, which may provide different roughness. Figure 2 shows some typical base constructions for $\Phi = 1.0$.

Other types of bases, i.e. flat plane (Set 9), and generation approaches, i.e. ordered packing (Set 8) are also presented for reference. Only monolayer ordered bases are considered for 2D simulations (Set 11).

### C. Basal effect on velocity profile

There are three categories of velocity profile observed in all simulations. In the following, the results of Sets 2, 4, 6 are discussed as examples (Fig. 3). The average velocity, $v_x$, is normalized by the square root of $gd_p$, where $g$ is gravitational acceleration. The elevation $y$ is normalized by $d_p$. Note that all simulations are performed with the same mono-disperse sample [Fig. 1(b)], and only base constructions are different (Fig. 2).

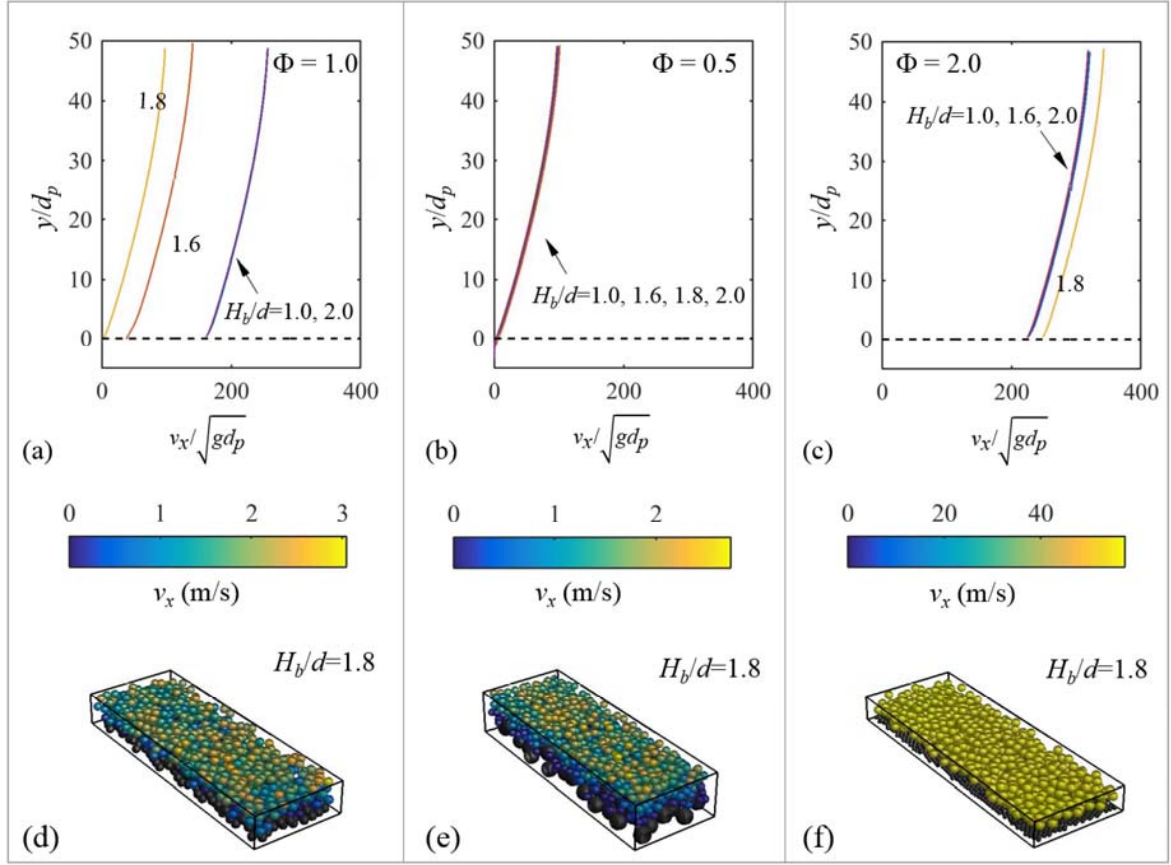

**FIG. 3 (color online). Basal effect. (a-c) Typical velocity profiles for $\Phi = 1.0, 0.5, 2.0$, respectively. (d-f) Flow snapshots near base (with $H_b = 1.8d$) for $\Phi = 1.0, 0.5, 2.0$, respectively.**

For $\Phi = 1.0$ (Set 4) in Fig. 3(a), the basal condition (thus velocity profile) is dependent on the base construction, i.e. the random spacing of the base surface. When the spacing is formed appropriately (e.g. by using $H_b = 1.8d$), sufficient roughness is obtained and no slip occurs at the base. When the spacing is either too large or small, considerable slip is observed. The snapshot taken near the base [Fig. 3(d)] shows that when $H_b = 1.8d$, some particles tend to be stopped or decelerated by the bumps. However, the voids are not so deep as to totally capture the flowing particles, which are still part of the entire flow.

For $\Phi = 0.5$ (Set 2) in Fig. 3(b), where the base particles are twice the size of flowing particles, all cases exhibit identical non-slip base conditions. The result is unaltered by base constructions. There is a small range of zero velocity a few particles away from the actually bottom, which is also seen in [10] when some large and deep voids exist among bumps. By checking the snapshot near the base [Fig. 3(e)], it is clear that this segment represents the range where particles are trapped in the voids of the base structure. In experimental studies where the flow velocity is measured based on side-view observation, the observable base velocity is non-zero as large base particles enclose the trapped ones. This is referred to as hole-filling mechanism in [11], where a critical size ratio for the maximum roughness is found in experiments. Further enlarging the base particles beyond the critical size does not promote the roughness, because the voids among large particles are filled by small particles.

For $\Phi = 2.0$ (Set 6) in Fig. 3(c), all cases exhibit similar behaviors and the flow is sliding significantly on the relatively small base particles. This observation is essentially unaffected by base constructions. The velocity profile is more steep, implying less inter-layer shearing. The snapshot [Fig. 3(f)] shows that the flowing particles can hardly intrude into the base structures. A reasonable deduction follows that the extreme case of a flat base (i.e. $\Phi \to \infty$) yields a plug flow where the velocity profile tends to be linear and free of gradient [10].

### D. Necessity to quantify roughness

In order to focus on the basal effect, we plot the velocity at the bottom of the sample, $v_{xb}$, for different base constructions for Sets 1–7 in Fig. 4. Base velocity, $v_{xb}$, is normalized by surface velocity, $v_{xs}$, to eliminate the sliding-induced velocity difference. The ratio can also be interpreted as the extent of shear propagation from the free surface to the base. Base construction, or its surface morphology, can be indicated by the normalized base thickness, $H_b/d$, in Fig. 4(a), or the packing density of base surface, $\eta_s$, in Fig. 4(b).

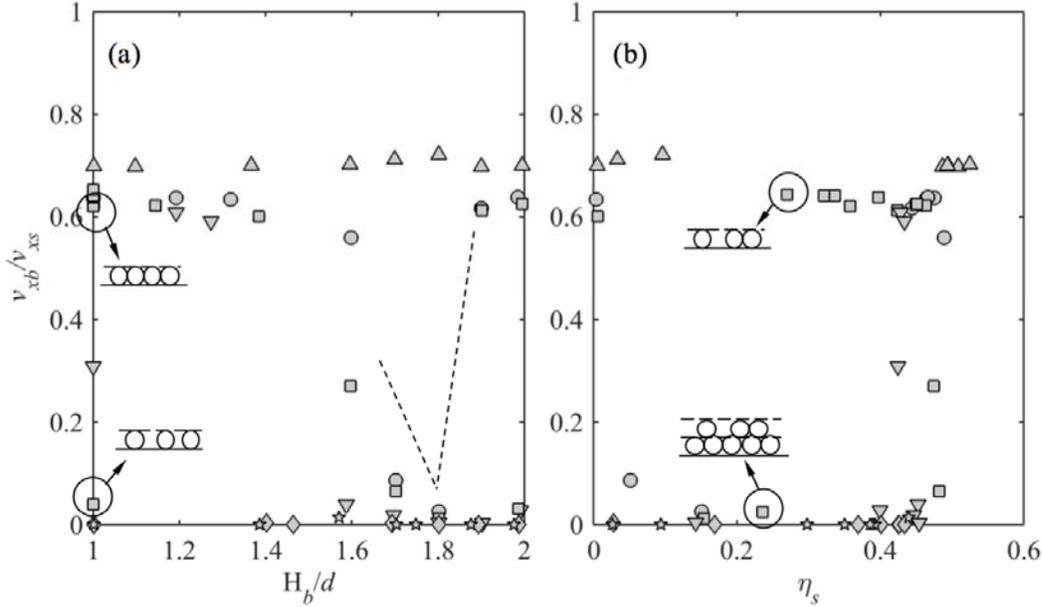

**FIG. 4.** Base velocity as a function of (a) base thickness and (b) packing density. Symbols △, ○, □, ▽, ◇, represent $\Phi = 2.0, 1.25, 1.0, 0.67, 0.5, 0.4$, respectively. Two dashed lines in (a) indicate the minima for $\Phi = 0.67, 1.0, 1.25$. Cases annotated by arrow-circles in (a) have the same base thickness but different packing density, and in (b) have similar base surface ($\eta_s$) but different overall base constructions.

Figure 4 shows that when $\Phi$ is either too large or too small, the base velocity is insensitive to different base constructions. In particular, when $\Phi = 2.0$, considerable slip ($v_{xb}/v_{xs} = \sim 0.7$) occurs regardless of $H_b/d$ and $\eta_s$, while $\Phi = 0.4, 0.5$ always provide non-slip basal condition. On the other hand, base construction procedures are influential at moderate size ratios. In Fig. 4(a), when $\Phi = 0.67, 1.0, 1.25$, $v_{xb}/v_{xs}$ is in general a function of $H_b/d$. A local minima corresponding to $H_b = 1.8d$ can be identified. The minimum indicates that an optimal roughness is achieved at a certain thickness (i.e. $\sim 1.8d$) when the base is generated by the method adopted in Sets 3–5 (Table I). Similarly, an optimal density around 0.2 is observed, corresponding to $H_b = 1.8d$. The optimization of base roughness associated with the thickness of base layers or the surface density will be elaborated in Sec. III.D.

Despite the clear minima around $H_b/d = 1.8$ and $\eta_s = 0.2$ for $\Phi = 0.67, 1.0, 1.25$, several issues are also revealed by Fig. 4. First, $v_{xb}/v_{xs}$ is not a single-variable function of either $H_b/d$ or $\eta_s$. Size ratio $\Phi$ must be specified to complement the description of basal velocity condition. Similarly, $\Phi$ is also not a sole indicator for base roughness. Second, non-unique basal conditions exist when $H_b = 1.0d$ and $2.0d$, an example being the annotated cases for $\Phi = 1.0$ (the lower case comes from Set 7). The non-uniqueness arises as $H_b$ poorly reflects the packing density within a certain thickness. When two bases have the same thickness but one is looser than the other, their base roughness can be much different [see sketches in Fig. 4(a)]. In contrast, packing density $\eta_s$ works around the second issue, but a third issue arises where similar packing density may yield quite different basal conditions [Fig. 4(b)]. An example is given by arrow-circle annotation. The two cases selected in Fig. 4(b) have the same packing density at the surface, but one surface is on top of a flat plane while the other on a dense layer. As the surface is loose enough ($\eta_s = \sim 0.28$) to expose the lower structure, the roughness contributed by the different lower layer gives rise to the discrepancy in velocity conditions. Indeed, a flat plane is generally smoother than a dense layer, hence the result in Fig. 4(b). However, contribution of the lower layer to base roughness has not been represented mathematically in previous studies. This issue reveals the need to quantify the roughness for a multi-layer base, taking into account contributions from different layers.

To summarize, a unique indicator is required to characterize the roughness of a base constructed by spheres, which should (1) simultaneously take into consideration size ratios and base constructions, and (2) consider bases as an assembly of multiple layers.

## III. CHARACTERIZATION OF BASE ROUGHNESS

### A. Local roughness

For clarity, we consider a layer of randomly packed spheres with the same size. The layer is partially filled and the centroids of all particles are coplanar [Fig. 5(a)]. The layer is equivalent to its projection on the plane accommodating all centroids, and the great circles representing all spheres on the projection plane do not overlap [Fig. 5(b)]. Next we use the Delaunay Triangulation (DT) scheme to discretize the plane into small triangular patches. The vertices of the triangles are the centers of the great circles, and the DT scheme ensures that no centroid is inside any discretized triangles. Now the roughness of the layer refers to the statistics of the local roughness considered at each individual DT triangle [Fig. 5(c)]. Note that in Fig. 5(b) ghost particles are placed outside the computational domain due to the use of periodic boundaries.

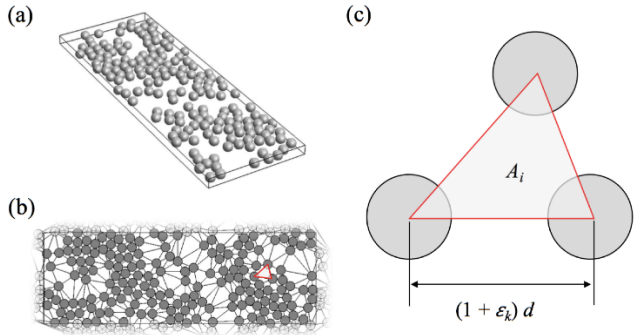

**FIG. 5.** Discretization of a surface made by spheres. (a) original layer, (b) projection and discretization, (c) a discretized triangle.

At an arbitrary triangle *i*, the spacing [5,11] refers to the shortest distances, subtracting the diameter of particles, between any two particles [Fig 1(a)]. It is denoted by $\varepsilon_k d$ with $k$ = 1, 2, 3 for the three sides in Fig. 5(c). Notice that the area occupied by particles in a triangle is invariant (i.e. half a circle's area). The area of the triangle, $A_i$, represents well the size of the local void. Both $\varepsilon_k$ and $A_i$ can be adopted to describe the spacing. In 2D, as the area is reduced to a length between two centers, $\varepsilon$ becomes the only quantity regarding the local spacing [5].

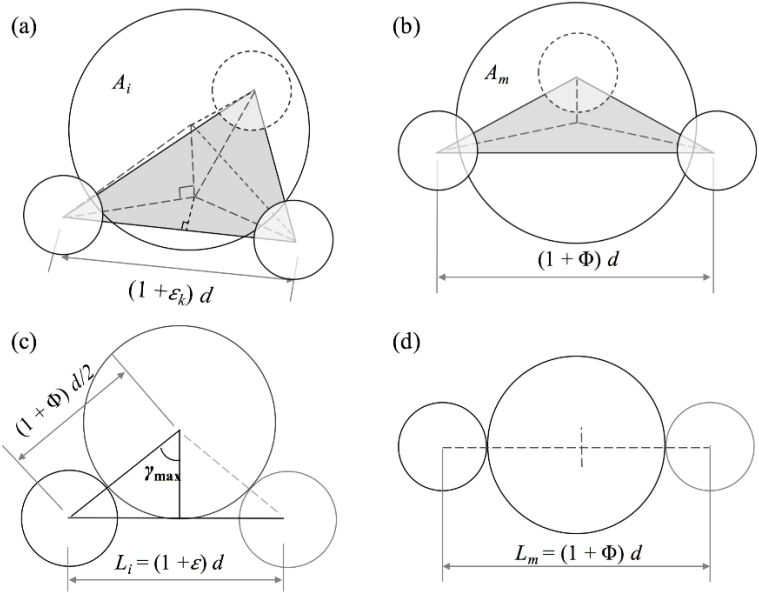

**FIG. 6. Definition of local roughness. (a) an arbitrary void in 3D, (b) the most stable situation in 3D, (c) an arbitrary void in 2D, (d) the most stable situation in 2D.**

Now base roughness can be interpreted as follows:

*If a sphere is placed onto the local triangular space under consideration, how stable would it be before it is mobilized by a tangential force?*

It is similar to the concept of static angle of stability [5,6,11]. In 3D, we consider the particle being commonly tangential to the three base particles composing an arbitrary triangular space [Fig. 6(a)]. The most stable situation [Fig. 6(b)] is where the void is so appropriate that the centroid of the placed particle is exactly coplanar with the DT triangle. If the void area at this most stable situation is $A_m$, the local roughness is defined by

$$R_{ai} = \frac{A_i}{A_m} \qquad (1)$$

where $R_{ai}$ is the roughness determined by area ratio, $A_i$ is the triangle area and *i* denotes the *i*th triangle. It is easy to prove that $A_m$ is the area of an equilateral triangle of side length $(1+\Phi)d$. For an arbitrary triangle, since the spacing is the most accessible measurement in both numerical and experimental situations, the area $A_i$ is formulated by the lengths of the three sides $(1+\varepsilon_k)d$ according to Hero's formula, $A_i = \sqrt{p \prod_k (p - a_k)}$, where $p = 1/2 \cdot \Sigma a_k$ is the semi-perimeter and $a_k = (1+\varepsilon_k)d$ is the side length of the triangle. Then we have

$$A_i = d^2 \sqrt{\frac{\sum(1+\varepsilon_k)}{2} \cdot \prod_k \left(\frac{\sum(1+\varepsilon_k)}{2} - (1+\varepsilon_k)\right)} \tag{2a}$$

$$A_m = \frac{\sqrt{3}}{4}(1+\Phi)^2 d^2 \tag{2b}$$

where $\varepsilon_k$ is the spacing at the three sides, $k = 1,2,3$ [Fig. 6(a) and Fig. 6(b)]. With Eq. (2a) and Eq. (2b), the definition of local roughness becomes

$$R_{ai} = \frac{\sqrt{\frac{\sum(1+\varepsilon_k)}{2} \cdot \prod_k \left(\frac{\sum(1+\varepsilon_k)}{2} - (1+\varepsilon_k)\right)}}{\frac{\sqrt{3}}{4}(1+\Phi)^2} \tag{3}$$

It can be seen that $R_{ai}$ is a function of $\varepsilon$ and $\Phi$, which considers both base construction and size ratio. Consistent with the observations in Sec. II, a wider spacing or smaller size ratio generally enhance the roughness. In 2D, since the area of a triangle is reduced to the length of a line, the roughness can be simplified into:

$$R_{ai} = \frac{L_i}{L_m} = \frac{1+\varepsilon}{1+\Phi} \tag{4}$$

where $L_i = (1+\varepsilon)d$ is the center-center length of void $i$ [Fig. 6(c)] and $L_m = (1+\Phi)d$ represents the most stable situation shown in Fig. 6(d). Interestingly, Eq. (4) is identical to the maximum contact angle [ $\gamma_{max}$ in Fig. 6(c)] described in [5], which extends the physical meaning of our definition. Dippel et al. [5] found that a single ball (in 2D) down a rough line can regularly impact with the fixed balls and reach a steady velocity at some inclinations. The average velocity is explicitly dependent on $\sin\gamma_{max}$ [i.e. $R_{ai}$ in Eq. (4)]. When $\sin\gamma_{max}$ is high, more tangential velocity is transferred to normal velocity during impacts, and the average velocity is lower. In other words, the base is rougher with a high $\sin\gamma_{max}$.

### B. Multi-layer composition

Before extending local roughness to global roughness, we consider the situation of a base constructed with multi-layer particles. In this case, the flowing particle can get in contact with the lower layer of the base, only if particles in upper layer of the base are not too densely packed, or in other words, the triangle area (void size) associated with the upper particles is larger than a critical area, $A_{cr}$. To illustrate the concept of $A_{cr}$, a reference plane in Fig. 7(a) [or straight line in 2D in Fig. 7(b)], lying beneath the three base particles, is used to represent either a flat plane or the top surface of the underneath layer. In the latter case, we assume the bumpy surface of the lower layer as flat for simplification. Some scenarios where this assumption may not be applicable is discussed in Sec. IV.D. Nevertheless, the inapplicability would only alter the formulation of $A_{cr}$, and not change the basic flow presented in this section. The determination of $A_{cr}$ is dependent on size ratio. When $\Phi \geq 1$, $A_{cr}$ is reached when the placed particle is commonly tangential to all three fixed particles and the reference plane. It can be solved as

$$A_{cr} = \frac{3\sqrt{3}}{4}\Phi d^2 \qquad (5)$$

as illustrated in Fig. 7(a). When $\Phi < 1$, the critical area is $A_{cr} = A_m = \frac{\sqrt{3}}{4}(1+\Phi)^2 d^2$.

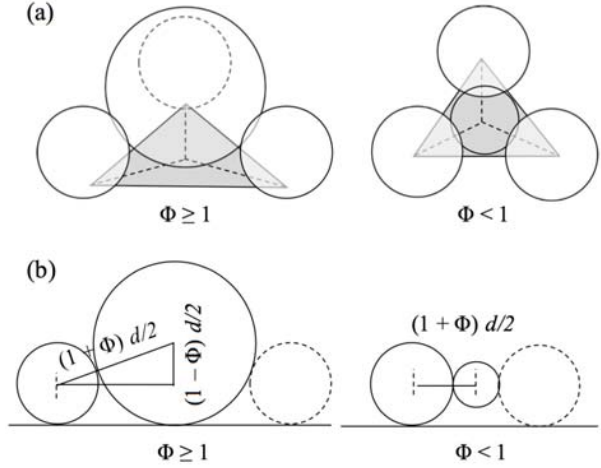

**FIG. 7. Critical void area in (a) 3D and (b) 2D situations.**

In 2D [Fig. 7(b)], the critical spacing is for $\Phi \geq 1$,

$$L_{cr} = 2\sqrt{\Phi} \cdot d \qquad (6)$$

and for $\Phi < 1$, $L_{cr} = L_m = (1+\Phi)d$.

The critical void size constitutes the basis of multi-layer composition. Figure 8 shows a general multi-layer situation where $N_l$ layers of particles are placed on top of a flat plane. The flat plane is counted as the 1st layer, as it is necessary in both experimental and numerical studies to serve as a substrate of the whole system. Typically, $N_l = 2$ or 3 when the base consists of one or two layer(s) of particles.

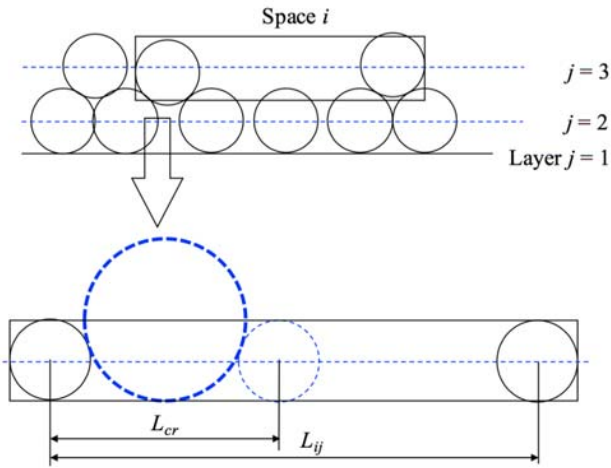

**FIG. 8. A base of multiple layers (2D for clarity). Replace $L$ as $A$ in 3D.**

Consider a void $i$ in layer $j$, whose area is $A_{ij}$, where $j$ is the layer number (Fig. 8). If $A_{ij} \leq A_{cr}$, the portion of layer $j–1$ beneath void $i$ contributes zero to the roughness. As void size increases, i.e. $A_{ij} > A_{cr}$, a portion of $(A_{ij}–A_{cr})/A_{ij}$ of layer $j–1$ beneath void $i$ is activated to provide roughness. An alternative interpretation is that layer $j–1$ is shielded by void $i$ at a percentage of area $A_{cr}/A_{ij}$, while the exposed portion of layer $j–1$ is $1–A_{cr}/A_{ij}$. Therefore, a weight function can be assigned to void $i$ at layer $j$,

$$w_{ij} = \min\left(\frac{A_{cr}}{A_{ij}}, 1\right) \tag{7a}$$

Here, the min function ensures $w_{ij} < 1$ if $A_{ij} > A_{cr}$ and $w_{ij} = 1$ if $A_{ij} \leq A_{cr}$. Correspondingly, the weight function for layer $j–1$ beneath void $i$ is $w_{i,j-1} = 1 - w_{ij}$. In conjugation with the weight function, local roughness, $R_{ai}$, is rewritten to $R_{aij}$ and corrected by $A_{cr}/A_m$, which represents the maximum roughness provided as the critical area is reached. Thus,

$$R_{aij} = \min\left(R_{ai}, \frac{A_{cr}}{A_m}\right) \tag{7b}$$

where $R_{ai}$ is calculated independently from Eq. (3), $R_{aij}$ is the local roughness of the $i$th void at the $j$th layer, and the additional subscript $j$ indicates that the min function is invoked due to the presence of layers [thus the critical area, Eq. (5)].

To compose the roughness of multiple layers, firstly the arithmetic mean of local roughness, $R_{aij}$, and local weight function, $w_{ij}$, is calculated over layer $j$,

$$R_{aj} = \frac{1}{N_j} \sum_{i=1}^{N_j} R_{aij} \tag{8a}$$

$$w_j = \frac{1}{N_j} \sum_{i=1}^{N_j} w_{ij} \tag{8b}$$

where $R_{aj}$ and $w_j$ are the mean roughness and weight function at layer $j$, respectively, $N_j$ is the number of triangles at layer $j$, the suffix $ij$ denotes the $i$th void at the $j$th layer, and $j = 2, \ldots, N_l$. When $j = 1$, i.e. the flat plane, $R_{aj} = 0$ and $w_j = 1$ are immediately taken. Next, a weighted average is performed to every adjacent two layers in sequence from the bottom two layers ($j = 2$) to the top ($j = N_l$),

$$R_a = w_j R_{aj} + w_{j-1} R_{a,j-1}, j = 2,\ldots,N_l \tag{9}$$

where $R_a$ is the roughness of the whole base layers. Note that Eq. (9) represents an algorithm in a loop manner; It is executed $N_l–1$ times as $j$ increases from 2 to $N_j$ and starting from the 2nd execution, $R_{a,j-1}$ is updated by the result of the previous execution. Equation (9) considers the combination of size ratio and base construction [Eq. (3,4)] and the composition of arbitrarily multiple layers [Eq. (7–9)]. Note that the formulation of Eq. (9) implies that layer-averaged roughness is representative of the entire layer, which is particularly reasonable for a flat plane or a densely packed layer.

The procedure of multi-layer composition is the same in 2D scenarios, except that the void area $A$ is substituted by the spacing length $L$ in Eq. (7).

## C. Range of possible values for $R_a$

The defined roughness, $R_a$, is a function of $\Phi$ and $\varepsilon$, and the range of possible values for $R_a$ at a given $\Phi$ is presented in Fig. 9. Generally, $R_a$ increases as $\Phi$ decreases or $\varepsilon$ increases. However, the increase of spacing, $\varepsilon$, would not infinitely enhance $R_a$ for a given $\Phi$. An upper bound (UB) for $R_a$ exists when local voids open up to the critical area $A_{cr}$, since beyond this point a non-zero weight is given to the lower, smoother, layers. The upper bound is dependent on size ratio. When $\Phi \geq 1$, the upper bound is $R_a \leq A_{cr}/A_m$, where $A_{cr}$ is given by Eq. (5). When $\Phi < 1$, the maximum roughness is reached as $A_{cr} = A_m$, hence $R_a \leq 1$. Up- and down-wards arrows in Fig. 9 indicate the tendency that as $\varepsilon$ increases, $R_a$ first increases to UB and then drops upon the triggering of weighted average [Eq. (7)].

On the other hand, the lower bound (LB) corresponds to the situation of close packing where no spacing exists among the three base particles, i.e. $\varepsilon = 0$ [Fig. 2(g) and Fig. 2(h)]. The lower bound is thus $R_{ai} \geq 1/(1+\Phi)$. However, it is noted that the void cannot be completely closed by spheres in three dimensions, and the minimum void left can only allow a particle with $\Phi_0 = 0.155$ to percolate. This indicates that when $\Phi$ is smaller than a small value $\Phi_0$, we have $R_a \equiv 1$, i.e. both upper and lower bounds are 1.0 (Fig. 9).

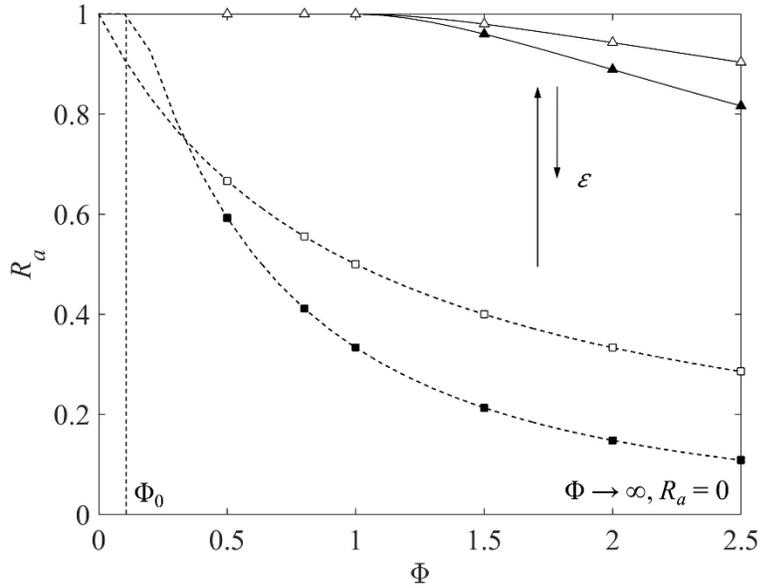

FIG. 9. Values of $R_a$ for different size ratio. Dash lines with filled and empty squares are lower bounds in 3D and 2D, respectively. Solid lines with filled and empty triangles are upper bounds in 3D and 2D, respectively.

Figure 9 also illustrates the scenario of a (frictionless) flat plane: with $\Phi \rightarrow \infty$, both UB and LB approach zero asymptotically. It indicates that a flat plane is the smoothest in terms of geometric roughness. Another source of roughness, namely, the microscopic contact friction, is discussed in Sec. IV.B.

The possible range of roughness is similar in the 2D scenario. For UB, applying Eq. (4) and Eq. (6) yields: when $\Phi \geq 1$, $R_{ai} \leq 2\sqrt{\Phi}/(1+\Phi)$; when $\Phi < 1$, $R_{ai} \leq 1$. For LB, since the void can be totally closed in 2D [Fig. 1(j)], $R_{ai} \geq 1/(1+\Phi)$ is simply applied. The LB and UB of $R_a$ in 2D are also visualized in Fig. 9.

## D. Optimal packing

As discussed above, the upper bound of $R_a$ is reached before the lower layers are activated since the lower layers (either flat plane or dense packing) are generally smoother. This indicates the existence of an optimal packing where the maximum roughness is reached for a given $\Phi$ (i.e. the UB lines in Fig. 9). Theoretically, the optimal packing is an equal-spaced triangular packing [e.g. Fig. 2(i)]. Each DT triangle is equilateral with area $A_{cr}$ [Eq. (5)].

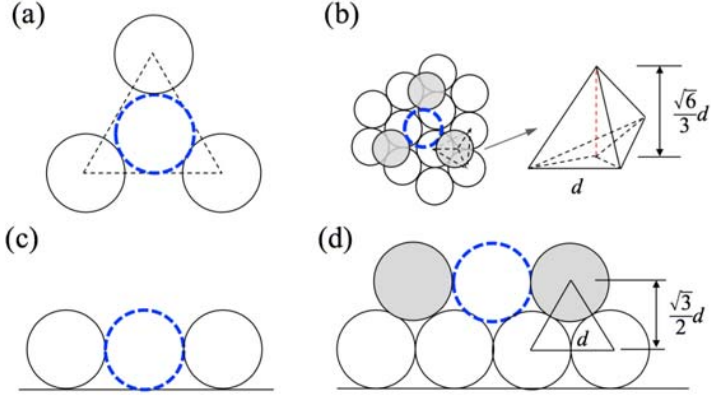

**FIG. 10. Typical optimal packing ($\Phi = 1$). Circles with solid and dashed borders are base and flow particles, respectively. Shadows distinguish base particles in different layers. (a) Plan view of a one-layer optimal base in 3D. (b) Plan view of a two-layer optimal base in 3D, where base thickness is around 1.82$d$. (c) Side view of a one-layer optimal base in 2D. (d) Side view of a two-layer optimal base in 2D, where base thickness is around 1.87$d$.**

Some equivalent expression of the optimal packing can be derived. For a base with one layer of particles [e.g. Fig. 10(a)], the optimal packing density (corresponding to critical area $A_{cr}$) is

$$\eta_{cr} = \frac{\pi}{9\sqrt{3}} \frac{1}{\Phi} \qquad (10)$$

which is applicable to arbitrary size ratios. When $\Phi = 1$, $\eta_{cr} = 0.2$. It confirms the observation in Fig. 4(b) that non-slip basal condition is generally achieved at surface density $\eta_s = \sim 0.2$, provided that base particles are large enough to provide roughness (i.e. $\Phi \leq 1.25$). For a base with two layers of particles, if the upper layer follows the optimal packing and the lower layer follows close packing [Fig. 10(b)], with $\Phi = 1$, it can be shown that the thickness of the base is $H_b = \sim 1.82d$. This is associated with the observation in Fig. 4(a), where the maximum roughness is achieved when $H_b = 1.8d$ is adopted in base generations. In 2D, the optimal packing density (by area) is ~0.39 for $\Phi = 1$, while the optimal base thickness is $1.87d$ in a two-layer situation. Despite the difficulties in constructing the optimal packing in practice, either in experimental or numerical studies, these typical values can be instructive to the generation of rough bases.

## E. Slip/non-slip condition

The new roughness, $R_a$, can indicate the slip/non-slip condition as a single variable. Figure 11(a) shows the data collected from 3D cases (Sets 1–7), in which three zones can be identified. In Zone I, where $R_a = 0.3$–$0.5$, sliding always occur near the base, where the size ratio can vary from 0.67 to 2.0. Ra = 0 represents the flat plane (Set 9). In Zone III, where $R_a \geq 0.62$, non-slip condition is always held, where the size ratio can range from 0.4 to 0.67, as long as the spacing is assigned appropriately. Therefore, a criterion of the imposition of non-slip condition is $R_a \geq 0.62$. Zone II represents a transition between Zone I and Zone III. A similar $R_a$ in this zone (where $\Phi =$ 0.67–1.25 only) may result in different basal conditions. A series of random seeds are adopted for Sets 3–5 to check repeatability. Error bars indicate more deviations in Zone II, the main source of error being the looser surface that allows more variations in base generation.

Similar phase diagram can be obtained for 2D cases (Set 11). Note that $\theta = 25°$ is used for 2D cases, which is close to the limit of steady flow in 2D [24,31]. As shown in Fig. 11(b), the transition is more abrupt in 2D, that is, no moderate basal velocity is observed in Zone II in the present cases. Another observation is that it is more difficult to achieve non-slip condition in 2D than in 3D. Two reasons are given; the first is that 2D voids capture beads less efficiently than 3D voids, given the same size ratio, and the second attributes it to the transversal motion of flowing beads near the base in 3D, which causes more energy dissipation [6].

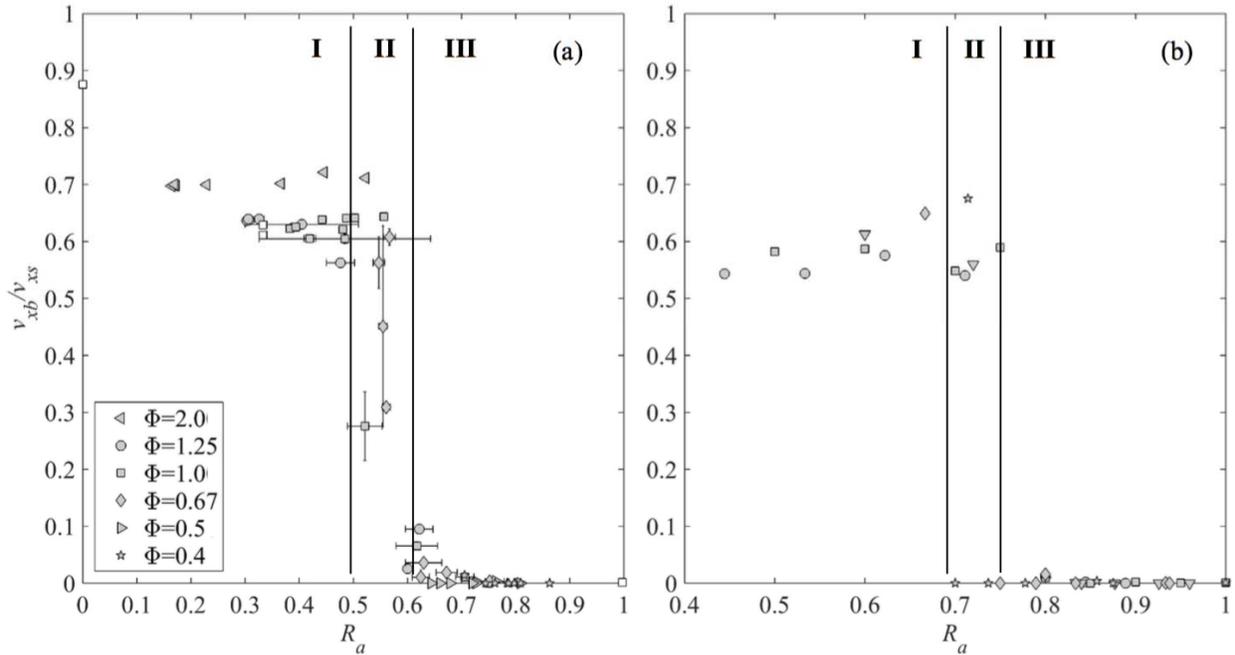

FIG. 11. Transition of slip/non-slip condition in (a) 3D ($\theta = 30°$) and (b) 2D ($\theta = 25°$). Error bars result from a series of different random seeds in simulations.

### F. Phase diagram

Angle of inclination is the driving factor for granular chute flows. It defines the kinetic energy gained from potential energy at SFD and the effective friction in the bulk of flows [1,2,8,24,25,31,44]. It is anticipated that the non-slip criterion at a higher inclination should also hold at lower inclinations. In Fig. 12, a series of different inclinations (i.e. $\theta = 20, 22, 25, 26, 28°$) is applied (Set 10). The overall trend is all similar in these cases, compared to the series with $\theta = 30°$. An intermediate range of $R_a$ is found connecting the zone of basal sliding ($R_a < 0.5$) and zero base velocity ($R_a > 0.62$). In the zone where non-slip condition is expected, the influence of $\theta$ is negligible. This proves the robustness of the criterion for non-slip rough base. In the zone where basal velocities are non-zero, $v_{xb}/v_{xs}$ is randomly influences by $\theta$. Indeed, the ac-tual magnitude of base velocity, $v_{xb}$, is a monotonic function of $\theta$, as shown in the insert graph of Fig. 12.

The optimal base thickness [Fig. 10(b)] for non-slip condition is also examined for different inclinations. The insert diagram of Fig. 12 presents the actual basal velocity as a function of base thickness for $\Phi = 1.0$ at different inclinations. The base velocity, $v_{xb}$, attains a minimum value of zero at $H_b/d = 1.8$, which is independent of the angle of inclination. The optimal thickness, $H_b = 1.8d$, is consistent with the theoretical treatment shown in Fig. 10(b). This once again proves that the imposition of non-slip condition is robust as long as an appropriate combination of size ratio and spacing is designated.

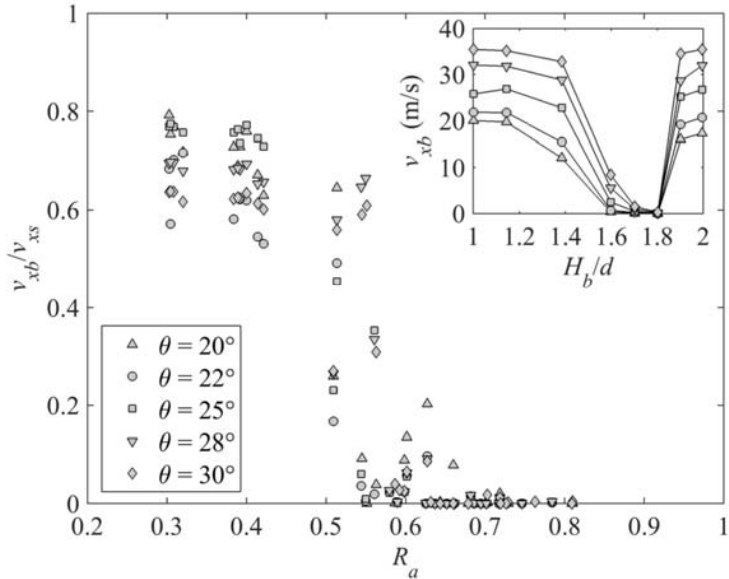

**FIG. 12. Slip/non-slip condition at different angles of inclination. Insert: basal velocity as a function of base thickness for $\Phi = 1.0$ at different inclinations.**

Now we further investigate the transition of slip/non-slip condition for different inclinations. Two boundaries can be identified (based on fitted trend-lines) for each $\theta$ to distinguish the three zones similar to Fig. 11(a). Lower inclinations are expected to enhance the chance of flow particles being trapped by the base, that is, the boundary should be generally left-shifted for lower $\theta$. A phase diagram can be established taking both inclination and base roughness into consideration (Fig. 13). For all inclinations under consideration, the boundary between slip and transition regime is consistently 0.51, whilst the boundary defines non-slip condition is inclined to slightly higher $R_a$ as $\theta$ increases.

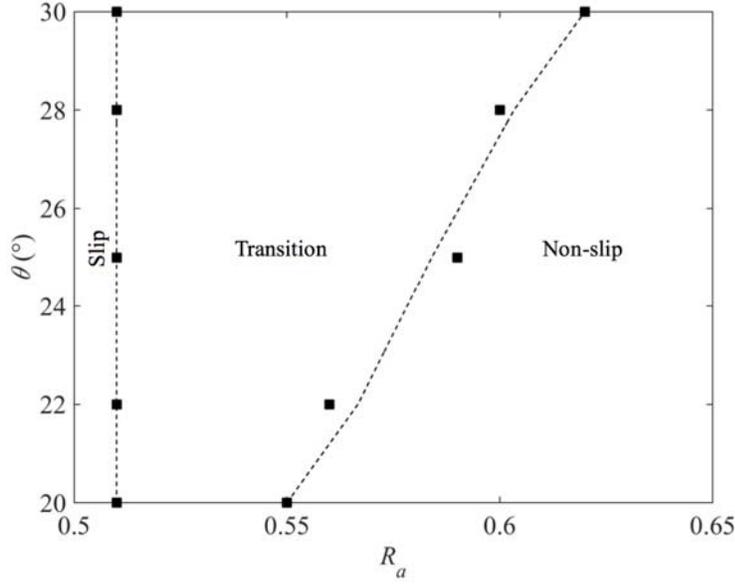

**FIG. 13.** Boundaries of slip/non-slip transition at different angles of inclination.

## IV. DISCUSSION

### A. Simplified definition

The definition of roughness is established step-by-step from a single triangle that compose a small piece of the base. This is mathematically rigorous, but may not be convenient in implementations. The same concept can be implemented in a simpler way, that is, to consider one layer as a whole instead of an assembly of discretized triangles. The packing density of each layer, which is easier to access in practice, can be used to derive a simplified roughness indicator. Let $R_p$ be the roughness defined by packing density. The equivalent void area at each layer, $A_j$, is derived from packing density, $\eta_j$, assuming equilateral triangulation,

$$A_j = \frac{\pi}{12\eta_j} d^2 \tag{11}$$

where $j$ is layer index. Weight function at layer $j$ is determined by $w_j = \min(A_{cr}/A_j, 1)$, while layer roughness, $R_{pj}$, becomes $R_p = A_j/A_m$. Multi-layer composition remains the same as Eq. (9). The simplified definition may lead to lower resolution in extreme cases where some local areas of the base surface are much denser than other areas. Nevertheless, for a randomly generated base, or a carefully-designed ordered base, the simplification is generally acceptable as examined in Fig. 14.

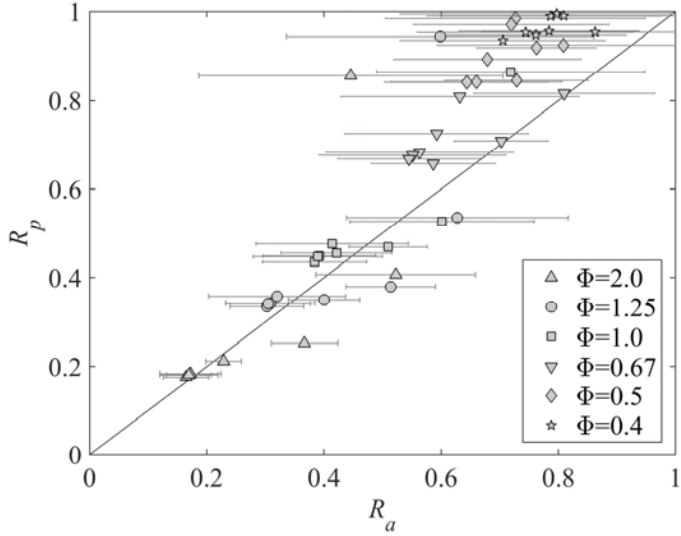

**FIG. 14.** Simplified definition vs. rigorous definition. Horizontal bars represent the range of local roughness, $R_{ai}$, before layer-average is performed.

Figure 14 shows the comparison between the roughness obtained by considering all individual local triangles ($R_a$) and the roughness obtained by considering only the layers ($R_p$), for Sets 1–6. It can be seen that, in general, the data points are lying near the 1:1 function, which goes through most horizontal bars. The width of horizontal bars indicates the standard deviation of local roughness, $R_{aj}$, obtained when averaging [Eq. (8a)] is performed for layer $j$. It represents rather the range of deviations than errors. In fact, this is the advantage of $R_a$ that by considering each individual DT triangles, a more rigorous description of the layer can be achieved. Note that in Fig. 14, more data points appear above the 1:1 line, meaning $R_p$ tends to overestimate the base roughness. This is attributed to the equilateral assumption made in Eq. (11), as an equilateral triangle has the greatest area in all triangles with the same perimeter. For the purpose of practical use, though, simply calculating $R_p$ is still recommended as the agreement with $R_a$ is generally good. Based on the phase diagram constructed by $R_p$ (not shown), the criterion for non-slip condition is $R_p > 0.8$.

## B. Contact parameters

In all previously presented cases, the contact parameters ($\mu$ and $e$) between the base and the flow are set to be 0.5. The roughness we have discussed so far is mainly the "geometrical roughness", which is a combination of the morphology of the base surface and the size ratio between flow and base particles. In a more general sense, roughness also depends on the coefficient of friction, which constitutes the Coulomb's friction law. It is expected that when the geometrical roughness is sufficient, it becomes the dominant factor over the coefficient of friction. In order to investigate this, the coefficient of friction, $\mu$, between the flowing particle and the base particles is varied from 0.1 to 1.0. Note that the contact parameter among bulk particles are not altered, otherwise the overall flow velocity will not be comparable; it has been revealed that velocity profile is significantly affected by the inter-particle friction, $\mu_p$, in bulk particles [10,24].

We define the differences caused by the change of $\mu$ as the standard deviation of basal velocity as $\mu = 0.1, 0.5, 1.0$, viz.

$$\sigma_\mu = \sqrt{\frac{1}{N_\mu}\sum_{i=1}^{N_\mu}\left(v_i - \bar{v}\right)^2} \quad (12)$$

where $\sigma_\mu$ is the standard deviation with respect to $\mu$, $v_i = v_{xb}/v_{xs}$ is the normalized basal velocity for the $i$th case, with the mean being $\bar{v}$, and $N_\mu$ is the number of different $\mu$, here $N_\mu = 3$.

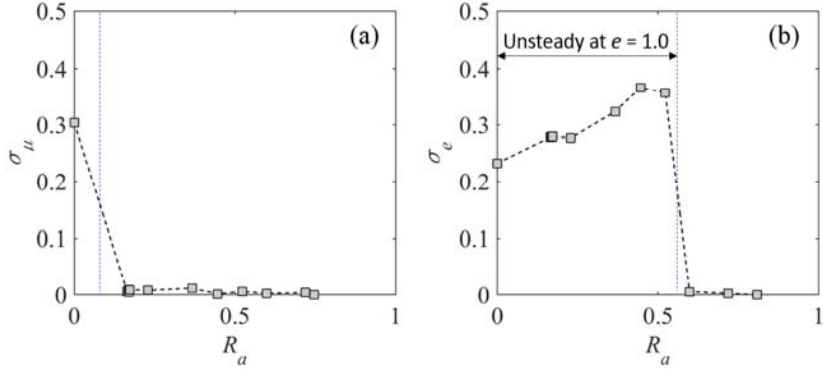

**FIG. 15.** Effect of (a) $\mu$ and (b) $e$ as a function of $R_a$. Vertical dash lines indicate transition.

In Fig. 15(a), $\sigma_\mu$ is plotted as a function of $R_a$ for selected cases in Sets 1–7, 9 with varied $\mu$. It can be seen that $\mu$ plays a primary role when the base is made by a flat plane ($R_a = 0$). In fact, it is verified that plug flows occur when $\mu = 0.1$, and a major change of flow structure is observed where flowing particles are arranged on an ordered lattice. This is referred to as crystallization and has been studied in some previous work [10,17,18]. When $\mu = 0.1$ and 1.0, the flow regime is normal. As $R_a$ increases to 0.3, the variation induced by $\mu$ drops rapidly to nearly zero. All cases with $R_a \geq 0.3$ are not influenced by the choice of $\mu$. This implies the existence of a transition where geometrical roughness (i.e. macroscopic friction) starts to dominate the roughness of a surface. The transition takes place at very small values of $R_a$, meaning that even a small amplitude of bumps ($\Phi \leq 2$) on the base will significantly hinder the effect of inter-particle friction (i.e. microscopic friction). A hypothesis proposed herein is that the tangential contact has little effect for a base to prevent sliding and capture the bottom of flowing particles. Indeed, in a fully developed granular flow, the apparent (inter-layer) friction, $\mu(I)$, is only a function of inertial number, I, or a function of the angle of inclination ($\tan\theta$), as reported in extensive studies [12,13,24,30,33]. It should be noted that tangential damping is not allowed in this study.

In contrast, we hypothesize that normal contact may play a more important role in the determination of basal conditions. To confirm this, we change the way energy dissipates at impact by varying $e$ from 0.1 to 1.0. The property of flowing particles is not varied. A similar indicator of the difference induced by $e$ can be defined as $\sigma_e$,

$$\sigma_e = \sqrt{\frac{1}{N_e}\sum_{i=1}^{N_e}\left(v_i - \bar{v}\right)^2} \quad (13)$$

where notations are similar with Eq. (12). Figure 15(b) presents $\sigma_e$ as a function of $R_a$. The effect of $e$ is significant at $R_a \leq 0.5$, which corresponds to the base made by flat plane or small beads. When $R_a$ is small, the contact angle (with respect to vertical) at normal impact is also small [5,6], and the dissipation of flow velocity is more dependent on $e$. The variation of $e$ from 0.1 to 1.0 retains the nature of slip condition, but substantially alters the magnitude of basal velocity. In this range ($R_a \leq 0.5$), sliding still occurs but in a gentler manner when $e = 0.1$, while when $e = 1.0$, the flows exhibit a plug-like, unsteady, pattern where the velocity profile is nearly vertical. Despite the high magnitude of $\sigma_e$ at $R_a \leq 0.5$, the increase of $\sigma_e$ in this range seems arbitrary because the flows turn unsteady as $e$ increases to 1.0. An abrupt transition is observed at $R_a = \sim 0.6$, beyond which the effect of $e$ is negligible [Fig. 15(b)]. In this case, geometric roughness ($R_a$) dominates over normal damping ($e$), and it is almost coincident with the transition from slip to non-slip condition ($R_a > 0.62$) as shown in Fig. 11(a). It indicates that the imposition of non-slip condition is independent of contact parameter (both $\mu$ and $e$), as the geometric aspects (i.e. the size of bumps and spacing) are more effective in preventing nearby particles from flowing. As $R_a$ increases, the maximum impact angle between flowing and fixed particles, $\gamma_{max}$, also increases ($R_a$ and $\gamma_{max}$ are equivalent in 2D [5]), which allows more tangential velocity to be transferred to normal velocity at impact. This transfer process is independent of $e$, provided that $R_a$ is sufficiently large (i.e. $R_a > \sim 0.6$).

The study on contact parameters ($\mu$ and $e$) shows that our major findings on $R_a$ and slip/non-slip condition with $\mu = 0.5$ and $e = 0.5$ are representative. While the frictional property of base materials ($\mu$) has little effect on the results except for a flat planar base, the damping parameter ($e$) is found crucial in determining the order or disorder regime of the flow (only when $R_a \leq 0.5$). In particular, if $e$ is high, the flow structure is easily crystallized and unsteady plug-flow is resulted. However, when $R_a > \sim 0.6$ the effect of $e$ vanishes, and thus the criterion of non-slip condition (i.e. $R_a \geq 0.62$) remains unchanged. A key implication is that geometric roughness plays an essential role in the basal effect of dense granular flows over bumpy bases, and its working mechanism is more about normal impact instead of tangential contact.

## C. Packing orientation

The orientation of DT triangles may also be a factor of roughness if a certain packing pattern is followed. For instance, Silbert *et al*. [10] presents two different perfectly ordered base (denoted as POB1 and POB2) where $\Phi = 1.0$, $\varepsilon = 0$. The packing orientation is in stream-wise (POB1) or span-wise (POB2) direction. It is found that POB1 is generally smoother than POB2, although both cases have $R_a = 0.58$ according to Eq. (3). We reproduce the two cases as Fig. 2(g) and Fig. 2(h), respectively. For a variety of slope angles ($\theta = 20–30°$), the flows on the stream-wise ordered base are generally faster than those on the span-wise ordered base (Fig. 16). Although the difference of base velocities is less than 10%, the orientation of DT triangles indeed gives rise to a variation of basal condition. This difference caused by packing orientation is omitted in our definition when layer average [Eq. (8)] is performed. As a result, the same $R_a$ may refer to different situations distinguished by the flow direction. This indicates that for an ordered surface made by spheres of triangular packing, the surface roughness is anisotropic. Flows in one direction may be easier than in other directions. However, in the current context, the difference is subtle (Fig. 16), and furthermore, the packing orientation follows no specific pattern in most randomly-generated bases.

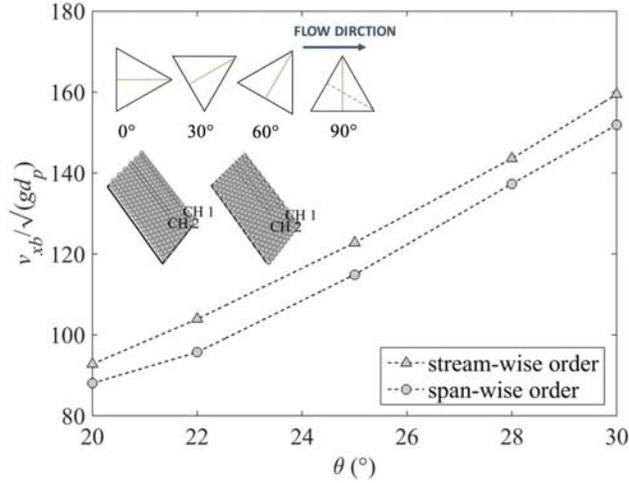

**FIG. 16. Effect of packing orientation ($\Phi = 1.0$, $\varepsilon = 0$; $R_a = 0.58$). Stream-wise order [also Fig. 2(g)] is made of triangles at 30° and 90°, while span-wise order [also Fig. 2(h)] is made of triangles at 0° and 60°.**

As a discussion, we point out that to factorize the effect of packing orientation is non-trivial. One way to do so is to consider the channels (e.g. CH1 and CH2 in Fig. 16) that connect the voids in the flow direction (Fig. 16). Let $\beta$ be the packing orientation with respect to the flow direction. When $\beta = 0°$ and 60°, which corresponds to span-wise order, a flowing particle needs to climb up the top of fixed particles in order to move from one void to the adjacent void following the channel. The potential of the inherent roughness is fully invoked, thus a factor of 1.0. On the other hand, when $\beta = 30°$ and 90°, the minimal elevation a ball needs to escape from one void is characterized by a factor $\cos 30° = 0.866$. However, the factor of 0.866 is less than the difference found in Fig. 16 (~0.95). More discussion is needed if packing orientation is the major concern in some context.

### D. Other applications

The roughness defined in this study is based on planar surfaces made by mono-disperse spheres. It is of general significance as long as the bump size is comparable to flowing particles. It can be readily extended to a variety of situations, with examples including a base of semi-sphere/circle, triangular asperities, staggered packing, and circular/curved surface. Fig. 17 provides some of the instances in 2D. Extension to three dimensions is straightforward. Although in reality, the elementary grains that compose a base may be more complicated in geometry (e.g. sands, crushed glass beads, or imperfect spheres), the concept of combining asperity size and spacing to quantify roughness is expected to be instructive.

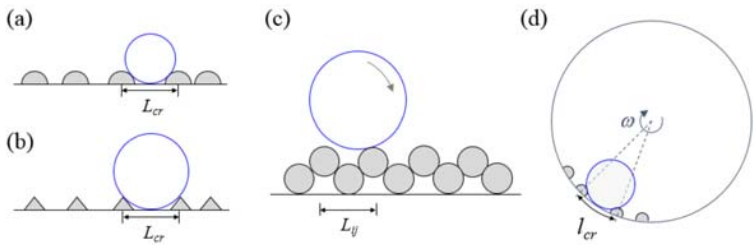

**FIG. 17. Bumpy bases made by (a) semi-circles, (b) triangular amplitudes, (c) circles on stagger lattice and (d) beads in a rotating drum. Notation $L_{cr}$ in (a) and (b) is the critical void length, $L_{ij}$ represents the spacing of a specific packing, and $l_{cr}$ in (d) is the critical arc length.**

In fact, the calculation procedure of $R_a$ is well compatible to the scenarios shown in Fig. 17. The determination of local roughness [Eq. (3) or Eq. (4)], the truncation by critical void area/length [Eq. (5) or (6), Eq. (7)] and the weight-average for multi-layer composition [Eq. (7–9)] remain the same. The major steps to be revised involve (i) how the void area/length is defined [for instance, arc length is used in Fig. 17(d)], and (ii) how the critical void area/length is derived. To illustrate the second point, generalization of the two scenarios in Fig. 17(a) and Fig. 17(c) are presented as examples below. Only two-dimensional configurations are considered for simplicity.

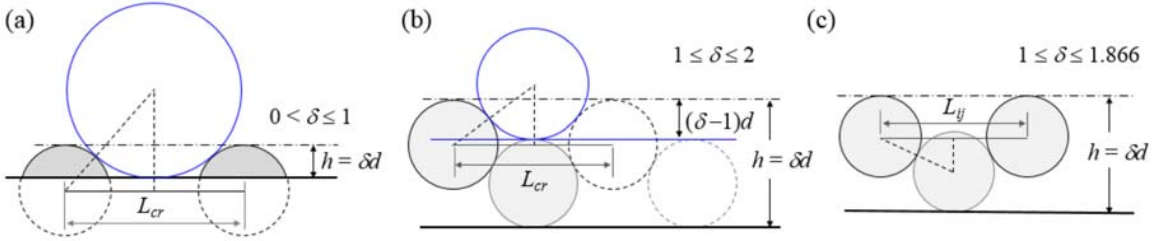

FIG. 18. Determination of $L_{cr}$ for (a) a line of arcs and (b) circles on a stagger lattice. (c) Spacing of a stagger layer of thickness $\delta d$.

The first configuration we present is a base made of equal-sized arcs [Fig. 18(a)]. The thickness of the base is given by $h = \delta d$, where $0 < \delta \leq 1$. Given a size ratio $\Phi$, it can be derived that the critical void length is

$$L_{cr} = 2d\sqrt{\delta\left[(1+\Phi)-\delta\right]} \tag{14}$$

and $L_{ij} = (1+\varepsilon)d$ remains unchanged [Eq. (4)]. When $\delta = 0.5$, which refers to the semi-circle base in Fig. 17(a), $L_{cr} = \sqrt{2\Phi+1}\cdot d$. If $\delta = 1.0$, it converts back to the normal base of a line of circles and $L_{cr} = 2\sqrt{\Phi}d$ is identical to Eq. (6).

The second example is a base consisting of circles on a stagger lattice [Fig. 17(c)]. It is of particular interest because in several previous studies of mono-disperse dense granular flows, the rough, bumpy base is constructed by duplicating a layer of bulk particles within a certain thickness, usually $1.2d$, and the layer generally follows the pattern of stagger ordering [10,41]. The duplication of bulk particles can provide acceptable non-slip condition in mono-dispersions (i.e. $\Phi = 1.0$). Here we make the thickness of the duplicated layer more general, i.e., $h = \delta d$ and $1 \leq \delta \leq 2$ [Fig. 18 (b)]. Note that the packing style (stagger layers) is different from the one adopted in Fig. 2(a–c), which places particles on top of a dense layer (little staggering). This difference underlies the different treatments of $L_{cr}$ between the two situations. In Fig. 18(b), a reference line (blue) is set on the top edge of the lower half layer, which compares to the reference line in Fig. 7(b). Considering the size ratio $\Phi$, it yields

$$L_{cr} = 2d\sqrt{(\delta-1)\left[(\Phi+1)-(\delta-1)\right]} \tag{15}$$

It can be examined that when $\delta = 2.0$, $L_{cr} = 2\sqrt{\Phi}d$ as it is reduced to the normal case. If $\delta = 1.2$ and $\Phi = 1.0$, which is widely used, $L_{cr} = 1.2d$. It is worth noting that if a stagger, regular, packing is followed [Fig. 18(c)], the spacing $L_{ij}$ can be specifically determined by $\delta$, as,

$$L_{ij} = 2d\sqrt{\delta(2-\delta)} \tag{16}$$

which is independent of Φ. In this case, $\delta$ is limited up to a close packing, $L_{ij} = d$, where $\delta \leq$ 1.866. For our particular interest, $\delta = 1.2$ yields $L_{ij} = 1.96d$, and the theoretical roughness can be evaluated following Eq. (4, 6–9) which gives $R_a = 0.6$. According to the phase diagram Fig. 13, $R_a = 0.6$ is adequate for non-slip condition at a wide range of inclinations, which confirms the rationale behind duplicating a flowing layer as the rough base for mono-disperse flows [10,41]. However, in the cases where size ratio between the base and flow particles is also a variable, the packing pattern of base should be more carefully designed, in which case the quantified roughness, $R_a$, will be much useful [20,36,42].

## V. CONCLUSION

In this paper, typical basal conditions encountered in granular chute flows are presented. It is found that the size ratio of flow/base particles and the construction of a bumpy base both determine whether basal slip occurs or not. A newly defined indicator of base roughness, $R_a$, can quantitatively consider both size ratio and base constructions to predict the slip/non-slip condition. It is generalized for random and regular packing of multi-layered spheres, in both two- and three-dimensional configurations. The transition of slip and non-slip condition is well indicated by $R_a$. For different inclinations, a phase graph is established showing the critical value of $R_a$, beyond which non-slip condition is respected. Some typical values of $R_a$, including its optimization, are provided for practical use. The extensions of the current definition to more general situations with different base geometries are also presented.

The presented base roughness results mainly from geometry. It is shown that geometric roughness is dominant over microscopic particle properties at contacts, including inter-particle friction and normal damping. While friction has little effect for most cases except for a flat plane, normal damping can profoundly influence the basal velocity and flow structure at low values of $R_a$ ($R_a <$ 0.62). When $R_a$ is higher, non-slip condition is achieved and the effect of normal damping vanish.

The characterization of base roughness presented in this work is of great value to future investigations of basal effect in granular flows. For instance, it can be applied to bi-disperse flows where the occurrence of size segregation may lead to a wide variety of base roughness.

ACKNOWLEGEMENT

The work was supported by Research Grants Council of Hong Kong (under RGC/GRF 17203614), the Research Institute for Sustainable Urban Development at The Hong Kong Polytechnic University, and FAP-DF, Brazil. The computation was performed using the HKU Information Technology Services research computing facilities that are supported in part by the Hong Kong UGC Special Equipment Grant (SEG HKU09).